\documentclass[sn-mathphys]{sn-jnl}
\jyear{2021}
\theoremstyle{thmstyleone}

\theoremstyle{thmstyletwo}

\theoremstyle{thmstylethree}

\begin{document}

\title[Article Title]{Zero field magnetic resonance spectroscopy based on Nitrogen-vacancy centers}

\author{\fnm{Linkai} \sur{Zhao}}
\author{\fnm{Q.} \sur{Chen}}
\affil{\orgdiv{Department of Physics and Key Laboratory of Low Dimensional Quantum Structures and Quantum Control of Ministry of Education}, \orgname{Hunan Normal University}, \orgaddress{\city{Changsha}, \postcode{410081}, \country{China}}}

\abstract{We propose a scheme to have zero field magnetic resonance spectroscopy based on a nitrogen-vacancy center and investigate the new applications in which magnetic bias field might disturb the system under investigation. Continual driving with circularly polarized microwave fields is used to selectively address one spin state. The proposed method is applied for single molecule spectroscopy, such as nuclear quadrupole resonance spectroscopy of a $^{11}$B nuclear spin and the detection of the distance of two hydrogen nuclei in a water molecule. Our work extends applications of NV centers as a nanoscale molecule spectroscopy in the zero field regime.}


\maketitle
\catcode`\|=12
\section{Introduction}\label{sec1}

Characterizing the properties of matter at the single-molecule level is of great significance in the development of science today, such as biology, chemistry, materials science. Nitrogen-vacancy (NV) center\cite{casola2018probing} is an emerging nanoscale quantum sensor\cite{2013quantumsenxor,fernandez2018sensing,abobeih2019atomic,london2013detecting}, which has important applications in quantum information\cite{2007register} and quantum metrology\cite{degen2017sening}. The NV centers have long coherence time at room temperature \cite{2003Longtime} and spin-dependent fluorescence properties, which can be optically initialized and read out\cite{ODMRgruber1997scanning,laser}. These excellent characteristics make NV can be used as a quantum sensor for detecting magnetic field\cite{wolf2015magnetic,maze2008nanoscale,hall2009sensing,schaffry2011proposed,balasubramanian2008nanoscale,taylor2008high}, electric field\cite{2011electricfiled}, temperature\cite{neumann2013temperature,acosta2010temperature,kucsko2013nanometre,toyli2013fluorescence} and strain\cite{knauer2020pressure}. NV-based nanoscale magnetometers is of great interest, which has been able to detect electron spins\cite{2015singleprotein}, nuclear spin-1/2 \cite{zhao2011atomic,cai2013diamond} and nuclear spins $I>1/2$\cite{lovchinsky2017magnetic,shin2014optically,henshaw2022nanoscale} in molecules, as well as detect a strongly coupled nuclear spin pair\cite{cai2013diamond,yang2018detection}.

Because traditional methods require a bias field\cite{2014biasfield} to lift the degeneracy of their ground state manifold. The bias field may disturb the system to be measured or even disrupting the measurement\cite{cai2013diamond,yang2018detection,zazvorka2020skyrmion}. For example, when the Zeeman effect induced by the magnetic field is stronger than the spin-spin coupling, it could be difficult for analyzing the molecular structure because crucial information of the chemical bond may be masked. Eliminating the dependence on bias field will help to miniaturize the instrument and further expand the application range of NV center\cite{2013appliactionrange}, such as using NV center to measure the spin system through magnetic couple interaction under zero-field conditions, and analyzing its energy level structure has natural advantages.

In recent years, several ways to control the NV center under zero-field are demonstrated \cite{vetter2022zero,2019zerofieldmagnetometry,wang2022zero,lenz2021magnetic,zhang2021robust,blanchard2007zero}. Zero field magnetic resonance spectroscopy based on Nitrogen-vacancy centers is of great interest. Here we propose a scheme for nanometer-scale nuclear resonance spectroscopy of the molecules lying on the diamond surface by using continual driving of circularly polarized microwave (MW) fields to control an NV center. We extend nuclear quadrupole resonance (NQR) studies of a $^{11}$B nuclear spin by using NV centers to the zero field regime. Also we applied our scheme to detect the distance of two hydrogen nuclei in a water molecule.

\section{The difficulty of using a conventional NV sensor at zero field and our method}\label{sec2}

\begin{figure}[h]
\includegraphics[width=3in]{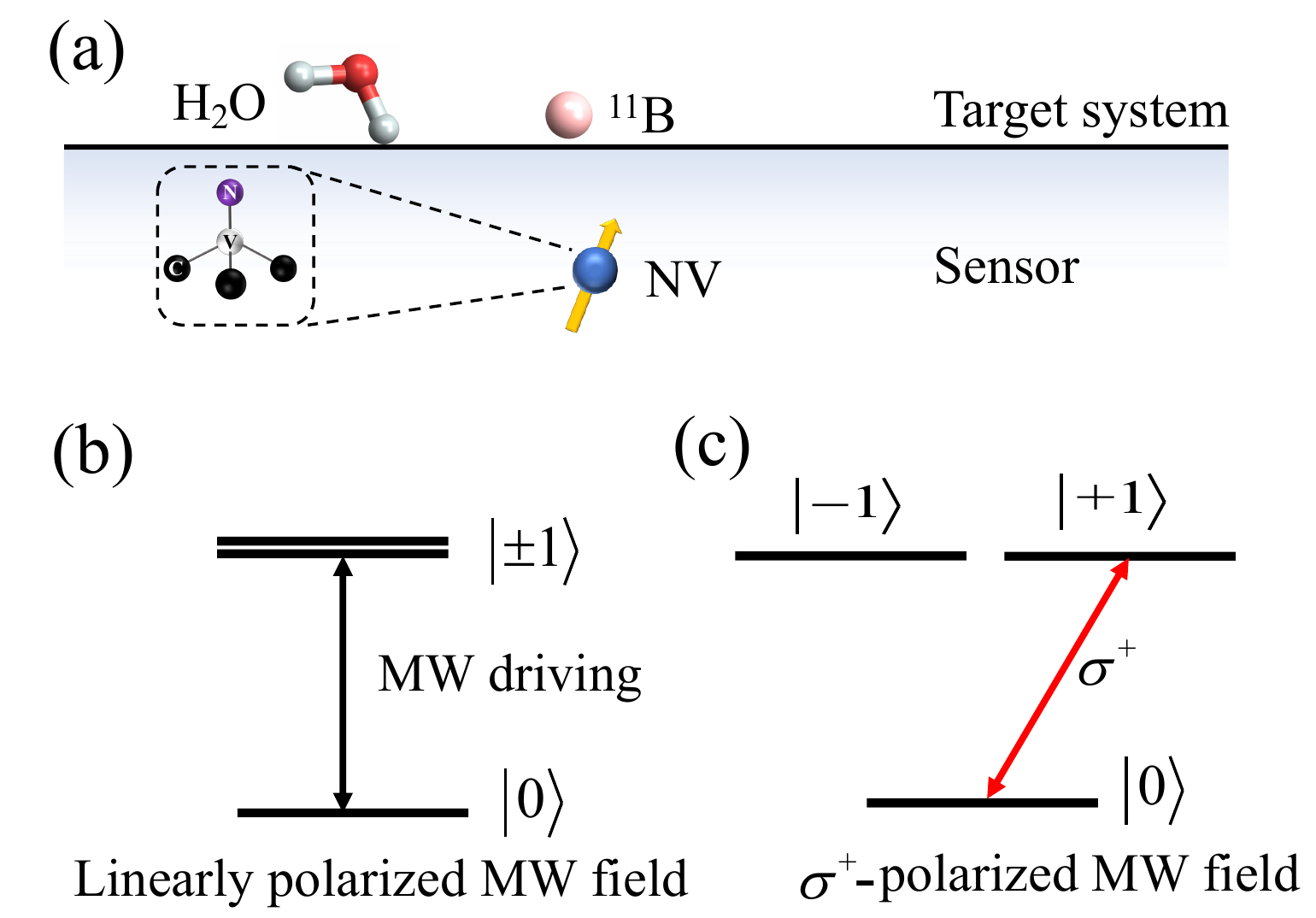}
\centering
\caption{(a) A shallow NV sensor is used for detecting target systems, i.e., a $^{11}$B nuclear spin or a water molecule on the diamond surface.  (b) A linearly polarized MW field is used for manipulation and the states $\left| +1 \right\rangle$ and $\left| -1 \right\rangle$ are degenerate at zero field. (c) The ${\sigma ^+\mbox{-}}$polarized MW field selectively drives the transition between $\left| 0 \right\rangle$ and $\left| { + 1} \right\rangle $ when the state $\left| -1 \right\rangle$ is not affected. }
\end{figure}

The conventional method uses an additional magnetic field to lift the degeneracy of  $\left|+1\right\rangle$ and $\left|-1\right\rangle$ and thus allows selective driving with a continuous microwave field of one specific electronic transition\cite{cai2013diamond}. By considering the NV center near the diamond surface, there exists magnetic dipole–dipole interaction between the NV spin and other spins in the target system on the diamond surface as shown in Fig. 1(a), when the MW driving matches a specific frequency in the target system, the state of the NV spin could be transferred. However, at zero-field, the energy levels between $\left| {{m_s} = \pm1} \right\rangle$ are degenerate, which is the main obstacle for the applications at zero field.

Suppose the linearly polarized MW field is given by ${H_w}=\sqrt2\Omega\cos\left({\omega t+\varphi}\right){S_x}$, where $\sqrt2\Omega$, $\omega$, $\varphi$ are the Rabi frequency, frequency and phase of the MW fields respectively. The corresponding Hamiltonian of the NV center reads as follows
\begin{equation}
H_{lin}=DS_{z}^{2}+\delta S_{z}+\sqrt{2}\Omega\cos\omega t{S_x},
\end{equation}
with a zero-field splitting $D=(2\pi)2.87$ GHz, $\varphi=0$ and $\delta$ is induced by an additional bias magnetic field. In the conventional method with an additional bias magnetic field, MW field can resonantly drive state transition between states $|0\rangle$ and $|+1\rangle$ when the state $|-1\rangle$ is not affected due to big energy mismatching. However, at the zero field, in the interaction picture with respect to ${H_0}=DS_{z}^{2}$ and $\omega=D$, we have
\begin{equation}
H'=\frac{\Omega}{\sqrt{2}}(|0\rangle\langle B|+|B\rangle\langle0|),
\end{equation}
in which $\left|B\right\rangle{\rm{=}}\frac{1}{{\sqrt2}}\left({\left|{{\rm{+}}1} \right\rangle + \left|-1\right\rangle } \right)$. So the linearly polarized MW field has the same effect to states $\left|+1\right\rangle$ and $\left|-1\right\rangle$, the linearly polarized MW field can resonantly drive state transition between states $|0\rangle$ and $|B\rangle$.

The magnetic dipole-dipole interaction between the NV spin and another spin gives with the secular approximation as
\begin{align}
H_{int}=-g{S_z}\left\{{3e^{z}_{r}( {e^{x}_{r}{I_x}+{e^{y}_{r}}{I_y}})+\left[{3{{\left({e^{z}_{r}} \right)}^2}-1}\right]{I_z}}\right\},
\end{align}
where $I_{i},i=x,y,z$ are the another spin operators, $g=\frac{{\mu_0}\hbar{\gamma_e}{\gamma}}{4\pi{r^3}}$ with  $r=|\vec{r}|$ denoting the distance from the NV spin to another spin. $\vec{e_{r}}=\vec{r}/r=(e^{x}_{r},e^{y}_{r},e^{z}_{r})$ is the unit vector connects the NV center and the target
spin, ${\gamma_e}$ and ${\gamma}$ are the gyromagenetic ratio of the electron spin and another spins, respectively.

When there is an additional bias magnetic field, one can take the NV sensor as an effective two-level system working in the subspace $\{|0\rangle,|+1\rangle\}$ by choosing a continuous microwave field with one specific frequency. However, when we take the NV sensor to detect a target spin at zero field, due to magnetic dipole-dipole interaction, although the linearly polarized MW field can resonantly drive state transition between states $|0\rangle$ and $|B\rangle$, it is difficult to treat the NV spin as an traditional effective two-level system interacting with the target system.

We solve this problem by applying circularly polarized MWs. Hamiltonians ${H_{w+}}$ and ${H_{w- }}$ describe the action of the ${\sigma^+\mbox{-}}$ and ${\sigma^-\mbox{-}}$ polarized MW field, respectively, which reads
\begin{align}
H_{w\pm}&=\frac{\Omega}{{\sqrt2}}\cos\left({\omega t}\right){S_x}\pm\frac{\Omega}{{\sqrt2}}\sin \left({\omega t}\right){S_y},
\end{align}
where $\frac{\Omega}{{\sqrt2}}$ is the Rabi frequency of the MW with frequency of $\omega=D$. In the interaction picture with respect to ${H_0}=DS_{z}^{2}$, the Hamiltonian is given by
\begin{align}
{H_{\pm}'}&=\frac{\Omega}{2}(|0\rangle\langle\pm1|+|\pm1\rangle\langle0|).
\end{align}
The effective Hamiltonian indicates that the electronic transition $|0\rangle\leftrightarrow|+1\rangle$ can be controlled by ${\sigma^+\mbox{-}}$polarized MWs, while the state $\left|-1\right\rangle$ remains unchanged. Similarly, one could also applied the ${\sigma^-\mbox{-} }$ polarized MW field to have state transition as $|0\rangle\leftrightarrow|-1\rangle$, when the state $|+1\rangle$ is not affected. We take ${\sigma^+\mbox{-}}$polarized MW field as an example, sketched in Fig. 1(c). Thus, we can also take the NV center as an effective two level system for quantum sensing.

\section{Detection of the nuclear spin}\label{sec3}
We first show the detection of a $^{11}$B nuclear spin $(I=3/2)$ which is on the diamond surface as an example. At zero field, the target system Hamiltonian is written as
\begin{align}
H_s=\frac{{\bar Q}}{{4I(2I-1)}}[3I_z^2-{I^2}+\eta(I_x^2-I_y^2)],
\end{align}
where the quadrupole coupling constant is $\bar Q=(2\pi)2.9921$  MHz. The quantity $\eta$ is the asymmetry parameter. The NV is placed at the origin of the coordinate system and the $^{11}$B nuclear spin is situated at position $\vec{r}$.

We apply ${\sigma^+\mbox{-}}$polarized MWs with frequency $\omega=D$ to drive the NV center. The Hamiltonian of the whole system is then
\begin{align}
H_w=&DS_{z}^{2}+\frac{\Omega}{{\sqrt2}}\cos\left({\omega t}\right){S_x}{\rm{+ }}\frac{\Omega}{{\sqrt2}}\sin\left({\omega t}\right){S_y}\nonumber\\
&+\frac{{\bar Q}}{{4I(2I-1)}}[3I_z^2-{I^2}+\eta(I_x^2-I_y^2)]\nonumber\\
&+{S_z}\left({{a_{x}}{I_{x}}+{a_{z}}{I_{z}}}\right),
\end{align}
where $\vec{I}$ is the spin-3/2 vector operator of the nuclear spin, ${{a_{x}}}$ and ${{a_{z}}}$ are the elements of the secular and pseudosecular hyperfine interactions, respectively, dependent upon the distance of the NV center and the nuclear spin. In the interaction picture with respect to ${H_0}=DS_z^2$,  the Hamiltonian is given by
\begin{align}
H^{\left(\pm\right)}=&\Omega \sigma _z+\frac{{\bar Q}}{{4I(2I-1)}}[3I_z^2-{I^2}+\eta (I_x^2-I_y^2)]-\frac{a_{z}}{2}I_z \nonumber\\
&+{\sigma_x}\left({{a_{x}}{I_{x}}+{a_{z}}{I_{z}}}\right),
\label{Eq9}
\end{align}
where ${\sigma_x}=\frac{1}{2}\left({\left|+\right\rangle\left\langle-\right|{\rm{+}}\left|- \right\rangle\left\langle+\right|} \right)$,$\sigma_{z}=\frac{1}{2}(|+\rangle\langle+|-|-\rangle\langle-|)$ with $\left|\pm  \right\rangle{\rm{=}}\frac{1}{{\sqrt2}}\left({\left|{{\rm{+}}1}\right\rangle\pm\left|0 \right\rangle}\right)$.

The target system eigenstates and the corresponding energies can be written as
\begin{align}
|\psi_1\rangle&=\frac{1}{{\sqrt {1{\rm{+}}{a^2}}}}\left({a{{\left|{+\frac{3}{2}}\right\rangle}_I}+{{\left|{-\frac{1}{2}}\right\rangle}_I}}\right),\\
{\lambda_1}&=\frac{1}{{48}}\left({15{\rm{+}}4\sqrt3\sqrt{3{\rm{+}}{\eta^2}}}\right)\bar Q,\\
|\psi_2\rangle&=\frac{1}{\sqrt{1{\rm{+}}{b^2}}}\left({{\left|{-\frac{3}{2}}\right\rangle}_I}-b{{\left|{+\frac{1}{2}}\right\rangle}_I}\right),\\
{\lambda _2}&=\frac{1}{{48}}\left({15{\rm{ + }}4\sqrt 3 \sqrt {3{\rm{ + }}{\eta ^2}} } \right)\bar Q,\\
|\psi_3\rangle&=\frac{1}{{\sqrt{1{\rm{+}}{b^2}}}}\left( {b{{\left|{+\frac{3}{2}}\right\rangle}_I}+{{\left|{-\frac{1}{2}} \right\rangle }_I}} \right),\\  {\lambda _3}&=\frac{1}{{48}}\left( {15{\rm{-}}4\sqrt 3 \sqrt {3{\rm{ + }}{\eta ^2}} } \right)\bar Q,\\
|\psi_4\rangle&=\frac{1}{{\sqrt{1{\rm{+}}{a^2}}}}\left({{\left|{-\frac{3}{2}}\right\rangle}_I}-a{{\left|{+ \frac{1}{2}}\right\rangle}_I}\right),\\
{\lambda_4}&=\frac{1}{{48}}\left({15{\rm{-}}4\sqrt3\sqrt{3{\rm{+}}{\eta^2}}}\right)\bar Q,
\end{align}
where $a=\frac{{\sqrt3+\sqrt{3+{\eta^2}}}}{\eta}$, $b=\frac{{\sqrt3-\sqrt{3+{\eta^2}}}}{\eta}$, and we assume $a_z\ll \bar Q$ is matched. One can find that the degenerate states $\left|{{\psi _1}}\right\rangle$ and  $\left|{{\psi _2}}\right\rangle$, and $\left|{{\psi _3}}\right\rangle$ and  $\left|{{\psi _4}}\right\rangle$. All the eigenstates and energies are determined by the interaction between the nuclear electric quadrupole moments and the local electric field gradients, which provides a way to know the electrostatic environment of the measured spins.

We choose the driving amplitude of the MW fields on resonance with the target nuclear spin to satisfy the Hartmann–Hahn matching condition which is given by
\begin{align}
\Omega=\frac{{\sqrt3\sqrt{3+{\eta^2}}}}{6}\bar Q.
\end{align}
The dipole-dipole interaction between the NV and nuclear spin will induce transition between the dressed states with only one transition frequency (non-zero frequency), which is determined by interaction between the nuclear electric quadrupole moments and the local electric field gradients $\bar Q$ and asymmetry parameter $\eta$.

In order to detect the target spin, initially NV center is prepared in state $\left|+\right\rangle $ and the nuclear spin is assumed in a maximally mixed state at room temperature. By considering that $a_z\ll \bar Q$, when the Hartmann–Hahn matching condition is matched, we have dominant flip-flop process and the flip-flip process is suppressed due to energy nonconservation. Assuming $\eta=0$, the probability of finding the dressed NV center, initially set to the state $\left| +\right\rangle $, in the state $\left|+\right\rangle $, after time t, is
\begin{align}
S(t)=1-\frac{{3a_{x}^2}}{{6a_{x}^2+8\delta{\Omega^2}}}{\sin^2}\left({\frac{{\sqrt {\frac{3}{4}a_{x}^2-\delta{\Omega^2}}}}{2}t}\right),
\end{align}
where $a_{x}=(2\pi)\times0.66$ kHz and $\delta\Omega=\Omega-\frac{{\sqrt3\sqrt{3+{\eta^2}}}}{6}\bar Q$. Therefore, it is quite similar to the case of detection of a nuclear spin spin-1/2 with a bias magnetic field\cite{cai2013diamond}. Although there is only one frequency and one cannot determine the two values $\bar Q$ and $\eta$ with a simple pure NQR measurement, the frequency in NQR has a maximum error of about 16\% in the determination $\bar Q$. That why normally NQR data are interpreted using the assumption $\eta=0$.

\begin{figure*}[htp]
\includegraphics[width=4.7in]{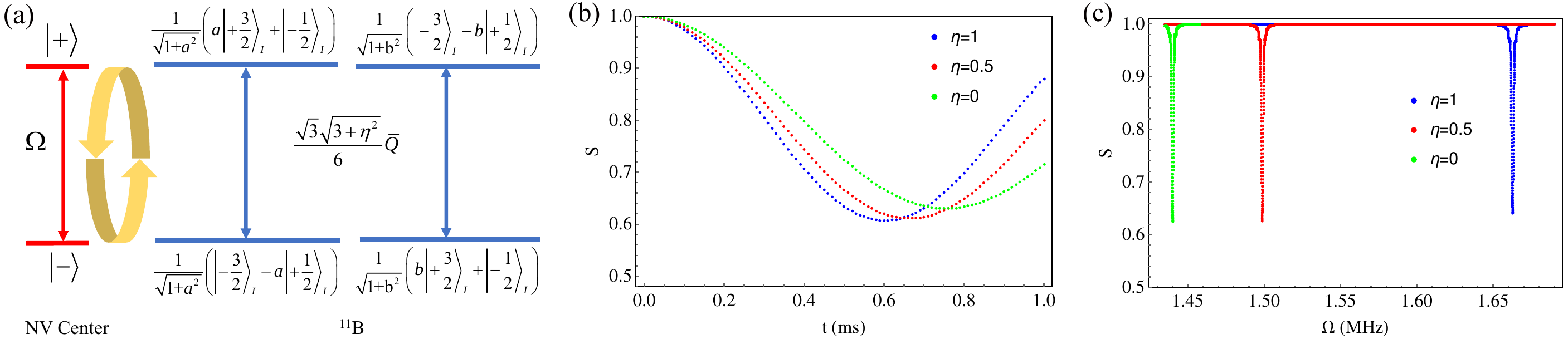}
\caption{$^{11}$B nuclear quadrupole resonance with an NV center at a distance of around 3 nm, and  we assume $a_{x}=(2\pi)\times0.66$ kHz. (a) When the Rabi frequency $\Omega$ is equal to $\sqrt3\sqrt{3+{\eta ^2}} \overline Q /6$, the flip-flop process will happen between the dressed NV spin and the target system, which leads to a change of the NV dressed population that can be measured via spin-dependent fluorescence of the NV spin. (b) The signal S (the population of state $|+\rangle$ of the NV spin) as a function of time t at resonance when $\eta$ varies (green: $\eta$=0, red: $\eta$=0.5, blue: $\eta$=1). (c) The signal S as a function of Rabi frequency $\Omega $ after evolution time t=0.75 ms.}
\end{figure*}

\section{Detection of the distance of two hydrogens in a water molecule}\label{sec4}
In this section, we consider two hydrogen nuclei in a water molecule on the diamond surface as our target system at zero field. The $^{1}$H$_{2}$O molecule is assumed to be around 5 nm from the NV center.

In the conventional method with an additional bias magnetic field, the system Hamiltonian of the target system is given by
\begin{align}
H_s'=\omega_{n}(I_{n1}^z+I_{n2}^z)+g_{12}'\left[{I_{n1}^zI_{n2}^z-\frac{1}{2}({I_{n1}^xI_{n2}^x+I_{n1}^yI_{n2}^y} )}\right],
\end{align}
where the spin operators $I^{x},I^{y},I^{z}$ are defined by magnetic field, $\omega_{n}$ is Larmor frequency of $^{1}$H, $g_{12}'=\frac{{\mu_0}\hbar\gamma_n^2}{4\pi d^3}(1-3\cos^{2}\theta)$, in which $\theta$ is the angle between the alignment of the hydrogen spin pair and the magnetic field. Furthermore ${\gamma_n}$ and $d$ denote the gyromagenetic ratio of the nuclear spin and the distance between two hydrogen nuclei. The target system eigenstates and the corresponding energies can be written as
\begin{align}
|E_0'\rangle&=\frac{1}{\sqrt {2}}({\left|{\uparrow\downarrow} \right\rangle+\left|{\downarrow\uparrow}\right\rangle}), &{E_0'}=&-\frac{g_{12}}{4}(1-3\cos^{2}\theta), \nonumber \\
|E_1'\rangle&=\frac{1}{\sqrt{2}}({\left|{\uparrow\downarrow} \right\rangle-\left|{\downarrow\uparrow}\right\rangle}), &E_1'=&0, \nonumber \\
|E_2'\rangle&=\left|{\downarrow\downarrow}\right\rangle,  &E_2'=&-\omega_{n}+\frac{g_{12}}{8}(1-3\cos^{2}\theta),\nonumber\\
|E_3'\rangle&=\left|{\uparrow\uparrow}\right\rangle,  &{E_3'}=&\omega_{n}+\frac{g_{12}}{8}(1-3\cos^{2}\theta),
\end{align}
where $g_{12}=\frac{{\mu_0}\hbar\gamma_n^2}{{2\pi d^3}}$. Eigenenergies are dependent on the angle between the alignment of the hydrogen spin pair and the magnetic field, and normally there is no degenerate states in the system. Because the alignment of the hydrogen spin pair is unknown and the obtained spectrum is complicated, one have to change magnetic field direction several times to determine the distance $d$\cite{cai2013diamond}. Under zero magnetic field, the target system eigenstates and the corresponding energies can be simplified as
\begin{align}
{|E_0\rangle}&=\frac{1}{\sqrt {2}} \left( {\left| { \uparrow  \downarrow } \right\rangle  + \left| { \downarrow  \uparrow } \right\rangle } \right), &{E_0}=&\frac{1}{2}{g_{12}}, \nonumber \\
{|E_1\rangle}&=\frac{1}{\sqrt {2}} \left( {\left| { \uparrow  \downarrow } \right\rangle  - \left| { \downarrow  \uparrow } \right\rangle } \right), &{E_1}=&0, \nonumber \\
{|E_2\rangle}&=\left| { \downarrow  \downarrow } \right\rangle,  &{E_2}=&- \frac{1}{4}{g_{12}}, \nonumber \\
{|E_3\rangle}&=\left| { \uparrow  \uparrow } \right\rangle,   &{E_3}=&- \frac{1}{4}{g_{12}}.
\end{align}

We assume the perpendicular coupling of the hydrogen spins to the NV center match $a_{1}^{z}$, $a_{2}^{z}\ll g_{12}$, all the eigenstates and energies are determined by the interaction between the two hydrogen nuclei, which provides a direct way to know the distance $d$ between two hydrogen nuclei. The corresponding energies are dependent upon the distance of two hydrogens. One can find that the states $|E_2\rangle$ and $|E_3\rangle$ are degenerate.

As shown in Fig. 3, the driving amplitude of the MW fields on resonance with the target system to satisfy the Hartmann–Hahn matching condition which is given by
\begin{align}
\Omega=\frac{3}{4}{g_{12}}.
\end{align}
The dipole-dipole interaction between the NV and nuclear spin will induce transition between the dressed states with only one transition frequency (non-zero frequency), which is determined by the distance between two hydrogen nuclei.

Initially NV center is prepared in state $\left|+\right\rangle $ and the two hydrogen nuclei are assumed in maximally mixed states at room temperature. By considering that $a_{1}^{z}$, $a_{2}^{z}\ll g_{12}$, when the Hartmann–Hahn matching condition is matched, we have the probability of finding the dressed NV center in the state $|+\rangle $ is
\begin{align}
S(t) = \frac{3}{4} + \frac{1}{4}\left[ {1 - \frac{{{a_x^2}}}{{{a_x^2} + \delta {\Omega ^2}}}{{\sin }^2}\left( {\frac{{\sqrt {{a_x^2} + \delta {\Omega ^2}} }}{2}t} \right)} \right],
\end{align}
where all the perpendicular and parallel components of interaction between the two hydrogen spins and the NV center are the same as $a_x=(2\pi)\times$0.63 kHz and $\delta\Omega=\Omega-\frac{3}{4}{g_{12}}$.

\begin{figure*}[htp]
\includegraphics[width=4.7in]{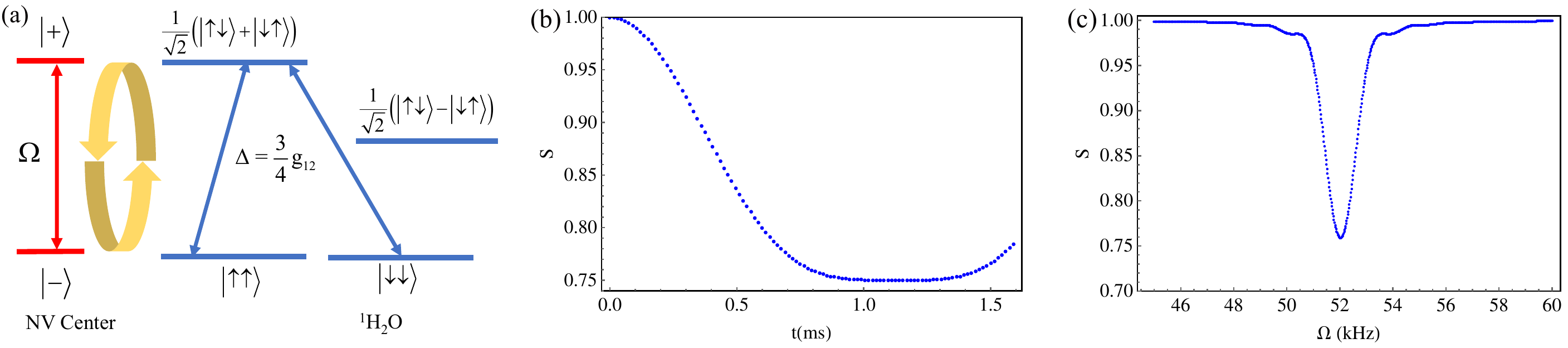}
\caption{Detection of the two hydrogens in a water molecule with an NV center at a distance of around 5 nm, and we assume $a_x=(2\pi)\times$0.63 kHz. (a) When the Rabi frequency $\Omega$ is equal to $\frac{3}{4}{g_{12}}$, the flip-flop process will happen between the dressed NV spin and the target system, which leads to a change of the NV dressed population that can be measured via spin-dependent fluorescence of the NV spin. (b) The signal S (the state $|+\rangle$ population of the NV spin) as a function of time t at resonance. (c) The signal S as a function of Rabi frequency $\Omega $ after time t=0.8 ms.}
\end{figure*}

\section{Summary}\label{sec4}
In conclusion, we have proposed a scheme to construct a nano-scale sensors based on NV centers in diamond at zero-field under MW continual driving. Continual driving with circularly polarized microwave fields is used to selectively address one spin state. The proposed device is applied for single molecule spectroscopy, such as nuclear quadrupole resonance spectroscopy of a $^{11}$B nuclear spin or the detection of the distance of two hydrogen nuclei in a water molecule. Our work extends applications of NV centers as a nanoscale molecule spectroscopy in the zero field regime.

{}
\end{document}